\documentclass[12pt,amsmath,amssymb]{article}
\pdfoutput=1
\usepackage{epsfig,amsmath,amssymb} 
\usepackage{graphicx}
\usepackage{dcolumn}
\usepackage{bm}
\usepackage{hyperref,cite}
\usepackage{color}
\usepackage{url,subcaption,comment}

\newcommand{\be}{\begin{equation} } 
\newcommand{\ee}{\end{equation} } 
\newcommand{\ba}{\begin{array} } 
\newcommand{\ea}{\end{array} } 
\newcommand{\bear}{\begin{eqnarray} } 
\newcommand{\eear}{\end{eqnarray} } 
 
\newcommand{\Tr}{{\rm Tr}}

\title{
\vspace*{-2.8cm}
\begin{flushright}
\normalsize{ Fermilab-PUB-17-401-T
  }
\end{flushright}
\vspace{.5cm}
\Large  
\textbf{Minimal $SU(3)\times SU(3)$ symmetry breaking patterns} \vspace*{0.2cm}   
}

\author{{\bf \normalsize Yang Bai$^a$ and Bogdan A. Dobrescu}$^{b}$
\vspace{5mm}
\\
$^a$\normalsize\emph{Department of Physics, University of Wisconsin-Madison, Madison, WI 53706, USA}  \vspace{1mm} \\
$^b$\normalsize\emph{Theoretical Physics Department, Fermilab, Batavia, IL 60510, USA}
}
\date{\normalsize October 4, 2017}

\begin{document}  
\setcounter{page}{0}  
\maketitle  

\vspace*{-0.3cm}  

\centerline{\bf Abstract}

\vspace*{0.1cm}  

{\normalsize
We study the vacua of an  $SU(3)\times SU(3)$-symmetric model with a bifundamental scalar. Structures of this type appear in various gauge theories such as the Renormalizable Coloron Model, which is an extension of QCD, or the Trinification extension of the electroweak group. In other contexts, such as chiral symmetry, $SU(3)\times SU(3)$ is a global symmetry. As opposed to more general $SU(N)\times SU(N)$ symmetric models, the $N=3$ case is special due to the presence of a trilinear scalar term in the potential. We find that the most general tree-level potential has only three types of minima: one that preserves the diagonal $SU(3)$ subgroup, one that is $SU(2)\times SU(2)\times U(1)$ symmetric, and a trivial one where the full symmetry remains unbroken. The phase diagram is complicated, with some regions  where there is 
 a unique minimum, and other regions where two minima coexist.
}
  
\thispagestyle{empty}  
  
\setcounter{page}{1}  
  
\tableofcontents  
  

\section{Introduction}
\label{sec:intro}

Several quantum field theories of interest for physics beyond the Standard Model have an $SU(3)\times SU(3)$ symmetry, which is spontaneously broken.
The embedding of the QCD gauge group, $SU(3)_c$, into an  $SU(3)_1\times SU(3)_2$ gauge symmetry 
has been considered in various contexts, including  dynamical symmetry breaking \cite{Preskill:1980mz}, rare $Z$ decays \cite{Hall:1985wz}, 
the study of heavy color-octet spin-1 particles such as the axigluon \cite{Frampton:1987dn, Buschmann:2017ucg} or the coloron \cite{Hill:1991at, Chivukula:1996yr}, composite Higgs models based on the top-seesaw mechanism \cite{Dobrescu:1997nm}, and so on. This requires the spontaneous breaking of the product group into its 
diagonal subgroup. A simple structure that achieves that breaking consists of  a
 single scalar field that transforms in the  bifundamental 
representation, with a potential that includes a trilinear interaction, as discussed in  the Renormalizable Coloron Model (ReCoM) \cite{Bai:2010dj, Chivukula:2013xka, Chivukula:2014rka}. A model of this type has been recently proposed as a solution to the Strong CP problem \cite{Kiel}.

\smallskip

The spontaneous breaking of  an $SU(3)\times SU(3)$ symmetry down to its diagonal $SU(3)$ group is also encountered
in certain tumbling theories \cite{Martin:1992aq}, latticized  extra dimensions \cite{Cheng:2001nh}, or the chiral symmetry of QCD with three light quark flavors \cite{Pisarski:1983ms}.  

\smallskip

Another example of symmetry breaking pattern is given by the so-called Trinification \cite{Glashow:1984gc, Achiman:1978rv, Babu:1985gi}, which is an embedding of the $SU(2)_W\times U(1)_Y$ 
electroweak group into an $SU(3)_L \times SU(3)_R$ gauge group. In that case the symmetry breaking may be achieved in two steps,
with the first one, $SU(3)_L \times SU(3)_R \to SU(2)_L \times SU(2)_R \times U(1)_{B-L}$, being again due to the vacuum expectation value (VEV) of a  bifundamental 
scalar. 

\smallskip

Here we study the scalar potential of the most general renormalizable potential for a scalar field that is an  $SU(3)\times SU(3)$ bifundamental.
Besides a mass term and two quartic terms, the potential includes a cubic term, or more precisely a trilinear interaction given by the determinant of the bifundamental,  which in $SU(N)\times SU(N)$-symmetric models  is 
specific only to the case $N=3$ (the determinant term is also present for $N=2$ or 4, but with a different mass dimension).
The parameter space spanned by the coefficients of these four terms leads to a nontrivial vacuum structure that has not been fully explored thus far.

\smallskip

Given the applications mentioned above, we are particularly interested in identifying the regions where the global minimum is either $SU(3)$ symmetric or $SU(2) \times SU(2) \times U(1)$-symmetric.
Also, we  would like to know if there exist vacua with other symmetry properties. In the absence of the cubic term in the potential, it has been known for a long time that there are 
no other nontrivial vacua \cite{Li:1973mq}.
In the presence of the cubic term, though, it is not immediately clear if other vacua exist.
For example, in Ref.~\cite{Glashow:1984gc} it is speculated that the potential for the
bifundamental scalar may have a minimum that preserves an $SU(2)\times U(1)$ group, and another minimum that breaks $SU(3) \times SU(3)$ down to $U(1)\times U(1)$.
We will prove that such patterns of symmetry breaking are not possible.

\medskip

Another question (partially addressed in \cite{Chivukula:2013xka}) 
is about the asymptotic behavior of the potential: what ranges of parameters make the potential bounded from below?
We find that two inequalities involving  the two quartic couplings are necessary and sufficient for that.

In Section \ref{sec:potential} we present the renormalizable potential and the parameter space. In Sections \ref{sec:su3-vacuum}-\ref{sec:less-vacuum} we identify all possible local minima.
The conditions for having a potential bounded from below are derived in Section \ref{sec:asymptotic}. We analyze the phase diagram of this theory, including all global minima, in Section \ref{sec:global-minimum}.
Section \ref{sec:coclusions} includes our conclusions.

\bigskip\bigskip

\section{$SU(3)\times SU(3)$ with a scalar bifundamental}
\label{sec:potential}
\setcounter{equation}{0}

Consider an  $SU(3)_1 \times SU(3)_2$  symmetry with a scalar $\Sigma$ transforming in the $(3, \bar{3})$ representation.  Thus, $\Sigma$ is a $3\times 3$ matrix with complex entries. 
The renormalizable potential of $\Sigma$ is given by 
\be
V(\Sigma) \,=\, -m_\Sigma^2\, \Tr(\Sigma \Sigma^\dagger) \, -\,  \left( \mu_\Sigma  \; {\rm det} \, \Sigma + {\rm H.c.} \right)  +
\frac{\lambda}{2}\left[ \Tr \left(\Sigma \Sigma^\dagger\right)  \right]^2 \,+\,\frac{\kappa}{2}\,\Tr \left(\Sigma \Sigma^\dagger \Sigma \Sigma^\dagger \right) ~.
\label{eq:sigma-pot}
\ee
The dimensionless couplings $\lambda$ and $\kappa$ are real numbers. The mass-squared parameter, $m^2_\Sigma$, may be positive or negative. The phase rotation freedom of $\Sigma$ allows us without loss of generality to choose the coefficient of the trilinear term (a 
mass parameter)  to be real and satisfy
\be
\mu_\Sigma \ge 0   ~~.
\ee
The potential $V(\Sigma)$ has an accidental $Z_3$ symmetry. If $\mu_\Sigma =0$, then the $Z_3$ symmetry is enhanced to a global $U(1)_\Sigma$ symmetry, with $\Sigma$ carrying nonzero global charge. We also note that when both $\mu_\Sigma=0$ and $\kappa=0$ the potential has an enhanced $SO(18)$ symmetry.

Even though the scalar $ \Sigma$ has 18 degrees of freedom,  
upon an $SU(3)_1 \times SU(3)_2$ transformation the most general form of its VEV is a diagonal $3\times 3 $ matrix. 
Furthermore, the diagonal $SU(3)_1 \times SU(3)_2$ transformations, associated with the  $T^3$ and $T^8$ generators,
can be used to get rid of two phases. Thus, the most general VEV of $ \Sigma$ has four real parameters:
\be
\langle \Sigma \rangle \, = \,  
{\rm diag} (s_1, s_2, s_3) \,  e^{i\alpha/3} ~~,  \;\; {\rm with}  \;\;   s_i \ge 0 , \; i = 1,2,3  \;  ,   \;\;\; {\rm and} \,\; - \pi <  \alpha \le  \pi   ~~~.
\label{eq:generalVEV}
\ee
The $1/3$ in the complex phase of the VEV is due to the $Z_3$ symmetry. We seek the values of $s_i$ and $\alpha$ that correspond to local minima of the potential. 

To identify the extrema of the $V(\Sigma)$ potential, we need to find $s_i, \; i = 1,2,3$ and $\alpha$ that satisfy the extremization (or more precisely stationarity) conditions, which are given by
\be
\frac{1}{2} \frac{\partial V } {\partial s_1} = \left( \lambda + \kappa \right)   s_1^3 +  \lambda  \, s_1  \left( s_2^2 + s_3^2 \right)  - \mu_\Sigma\, s_2 \,s_3 \, \cos  \alpha - m_\Sigma^2\, s_1 = 0  ~~,
\label{eq:first-deriv}
\ee
two analogous equations for $\partial V/ \partial s_2$ and  $\partial V/ \partial s_3$ (the $i = 1,2,3$ indices are cyclical), and finally
\be
\frac{\partial V } {\partial \alpha} =  2\,\mu_\Sigma \, s_1 \,s_2 \,s_3 \, \sin  \alpha  = 0  ~~.
\label{eq:phase}
\ee
This set of cubic equations in $s_i$ appears difficult to solve analytically; however, the first three equations can be replaced by a set of quadratic and linear equations as follows:
\bear 
&& \hspace*{-0.81cm}
\frac{\partial V } {\partial s_1} - \frac{\partial V } {\partial s_2}  = 2\left( s_1 - s_2 \right)  
\left[  \left( \lambda \!+\! \kappa \right)  \left( s_1^2 + s_2^2 + s_1 s_2  \right)   +  \lambda  \left( s_3^2 - s_1 s_2  \right)  +   \mu_\Sigma s_3 \cos  \alpha  - m_\Sigma^2    \right]  = 0
\nonumber \\ [2mm]
&&   \hspace*{-0.81cm}
\frac{\partial V } {\partial s_2} - \frac{\partial V } {\partial s_3}  = 2\left( s_2 - s_3 \right)  
\left[  \left( \lambda  \!+\!  \kappa \right)  \left( s_2^2 + s_3^2 + s_2 s_3  \right)   +  \lambda  \left( s_1^2 - s_2 s_3  \right)  +   \mu_\Sigma s_1 \cos  \alpha  - m_\Sigma^2    \right]  = 0
\nonumber \\ [2mm]
&&  \hspace*{-0.81cm}
 \frac{\partial V } {\partial s_2} + \frac{\partial V } {\partial s_3}  = 2\left( s_2 + s_3 \right)  
\left[  \left( \lambda  \!+\!  \kappa \right)  \left( s_2^2 + s_3^2 - s_2 s_3  \right)   +  \lambda  \left( s_1^2 +  s_2 s_3  \right)  -  \mu_\Sigma s_1 \cos  \alpha  - m_\Sigma^2    \right]  = 0 \,.
\nonumber \\  [-2mm]
&&  \hspace*{-0.81cm}
\label{eq:setDif}
\eear
To find the solutions to the set of equations (\ref{eq:phase}) and (\ref{eq:setDif}) we will consider a few separate cases.

A solution to the extremization conditions represents a local minimum if and only if the second derivative matrix has only positive eigenvalues.  Denoting that matrix by  $\partial^2 V/(\partial s_i \partial s_j)$ with $i, j = 1,..., 4$, where  $s_4 \equiv \alpha \,  \mu_\Sigma$, we find
\bear
&&  \hspace*{-0.68cm}  \frac{1}{2}
 \frac{\partial^2 V}{ \partial s_i \partial s_j } 
\nonumber  \\ [3mm]
&&  \hspace*{-0.68cm}  = 
  \left( \ba{cccc}  
(2 \lambda \!+ \! 3\kappa) s_1^2  + \Delta \;     &    \;    2 \lambda s_1 s_2 \!- \! \mu_\Sigma s_3  \cos\alpha    \;  &   \;   2 \lambda s_1 s_3 \!- \! \mu_\Sigma s_2  \cos\alpha    &    s_2 s_3  \sin  \alpha            \\ [3mm]
 \! \! 2 \lambda s_1 s_2  \!- \! \mu_\Sigma s_3  \cos\alpha   \;    &    \;    (2 \lambda  \!+ \! 3 \kappa) s_2^2   + \Delta     \;  &    \;    2 \lambda s_2 s_3 \!- \! \mu_\Sigma s_1  \cos\alpha  &    s_3 s_1  \sin  \alpha            \\  [3mm]
  \! \! 2 \lambda s_1 s_3 \!- \! \mu_\Sigma s_2  \cos\alpha   \;   &    \;    2 \lambda s_2 s_3 \!- \!\mu_\Sigma s_1   \cos\alpha    \;  &     \;   (2\lambda  \!+ \! 3\kappa) s_3^2 + \Delta      &     s_1 s_2  \sin  \alpha         \\ [3mm]
  s_2 s_3  \sin \alpha   &     s_3 s_1  \sin\alpha   &    s_1 s_2  \sin\alpha    &        s_1 s_2 s_3 \cos\alpha  \, /  \mu_\Sigma      \! \!
\ea \right) \,,
\nonumber \\ [-1mm]      \hspace*{-2.68cm} 
\label{eq:secondDerivative}
\eear
where we defined 
\be
\Delta \equiv  \lambda  \left(  s_1^2 + s_2^2  +  s_3^2 \right)  - m_\Sigma^2   ~~~.
\ee

Let us first apply these minimization conditions to the extrema located at the trivial solution to Eq.~(\ref{eq:setDif}), $s_1 = s_2 = s_3 = 0$, for any $\alpha$.
Three of the eigenvalues of $\partial^2 V/(2 \partial s_i \partial s_j)$ are equal to $- m_\Sigma^2$, while the fourth one is zero (representing a flat direction along $\alpha$).
Thus, there is a minimum with $V(\Sigma)=0$ at $s_1 = s_2 = s_3 = 0$ provided $m_\Sigma^2 < 0$.

\bigskip\bigskip

\section{$SU(3)$-symmetric vacuum}
\label{sec:su3-vacuum}
\setcounter{equation}{0}

We now search for minima that have $s_1 = s_2 = s_3 > 0$, so that the VEV preserves an  $SU(3)$ symmetry, which is the diagonal subgroup of 
the $SU(3)_1 \times SU(3)_2$  symmetry.
The three equations (\ref{eq:setDif}) are then replaced by a single quadratic equation:
\be
(3 \lambda + \kappa) s_1^2 =  \mu_\Sigma \, s_1 \cos\alpha + m^2_\Sigma ~~.
\ee
The extremization condition (\ref{eq:phase}) becomes $\sin\alpha = 0$. 
The phase $\alpha$ is further constrained by requiring 
stability of the potential.
The second-derivative matrix shown in Eq.~(\ref{eq:secondDerivative}) has an eigenvalue  equal to the 44 entry,
namely $s_1^3 \cos\alpha \, /  \mu_\Sigma$. Imposing that this is positive implies $\alpha = 0$.

For the range of parameters where 
\be
\mu^2_\Sigma > - 4(3 \lambda + \kappa) \, m^2_\Sigma   ~~,
\label{eq:mu-constraint}
\ee

\noindent
there are two solutions to the extremization conditions:
\be
s_1 =  s_2 = s_3 =  
\frac{1}{ 2 (3 \lambda + \kappa)} \left(
\pm \sqrt{4(3 \lambda + \kappa) \, m^2_\Sigma + \mu^2_\Sigma} \, + \mu_\Sigma \right) 
  ~~.
\label{eq:SU3minimum}
\ee
 Given that $s_i >0$,  the above solution with positive sign is valid only when
 \be
 3 \lambda + \kappa  > 0  ~~~,
 \label{eq:3lambda}
 \ee
while the  solution with negative sign requires  $m^2_\Sigma  < 0$.

We need to  determine the  regions of parameter space where these extrema satisfy the minimization conditions along the $s_i$ directions with $i = 1,2,3$. 
The $3\times 3$ upper-left  block of the second-derivative matrix shown in Eq.~(\ref{eq:secondDerivative}) may be written as follows:
\be
\hspace{-0.cm}
{\cal M}^2  = 
 \left( \ba{ccc}  
1 \;     &     0    &      0          \\ 
0  \;    &       1/\sqrt{2}     &      1/\sqrt{2}       \!\!       \\  
0  \;   &      -1/\sqrt{2}     &       1/\sqrt{2}     \!\!   
 \ea \right) 
  \frac{1}{2} \frac{\partial^2 V}{ \partial s_i \partial s_j }  
  \left( \ba{ccc}  \displaystyle
1 \;     &       0     &     0          \\ 
0  \;    &       1/\sqrt{2}      &     -1/\sqrt{2}   \!\!   \!\!        \\  
0  \;   &      1/\sqrt{2}     &      1/\sqrt{2}  
 \ea \right)   
 =
\left( \ba{ccc} 
 {\cal M}^2_{11}  \;     &        {\cal M}^2_{12}   \;  &      0          \\ [.2mm]
 {\cal M}^2_{12}   \;    &        {\cal M}^2_{22}     \;  &      0         \\  [.2mm]
0  \;   &    \;   0     &       {\cal M}^2_{3}    \!\!   
 \ea \right) , \\
 \label{eq:massSquared-matrix}
\ee
where the elements of the $2\times 2$ upper-left block of ${\cal M}^2$ are given by
\bear
& &  {\cal M}^2_{11} = \frac{  1 }{ 3 \lambda +  \kappa} \left[ 2 (\lambda +  \kappa ) \, m_\Sigma^2  +   (5\lambda + 3\kappa) \, \mu_\Sigma \, s_1  \right]  ~~,
\nonumber \\ [2mm]
& &  {\cal M}^2_{22} =   2 \, \frac{ 2 \lambda +  \kappa }{ 3 \lambda +  \kappa} \left(  m_\Sigma^2 + \mu_\Sigma \, s_1 \right)    ~~,
\nonumber \\ [2mm]
&&   {\cal M}^2_{12} =    \frac{ \sqrt{2} }{ 3 \lambda +  \kappa} \left[  2 \lambda \, m_\Sigma^2  - (\lambda + \kappa) \,\mu_\Sigma \, s_1   \right]     
   ~~,
\eear
and the 33 entry of ${\cal M}^2$ is 
\be
 {\cal M}^2_{3} =    \frac{ 2 }{ 3 \lambda +  \kappa} \left[  \kappa \, m_\Sigma^2  + (3\lambda + 2\kappa)  \,\mu_\Sigma \,  s_1   \right]      ~~.
\ee
The eigenvalues of ${\cal M}^2$ are the squared-masses of the radial modes. $SU(3)$ invariance implies that two eigenvalues  are equal,  $ {\cal M}^2_{2} =  {\cal M}^2_{3}$, because they 
are the squared-masses of different components (associated with the $T^3$ and $T^8$ generators) of an $SU(3)$-octet scalar.
The third eigenvalue represents the squared-mass of an $SU(3)$-singlet scalar, and is given by 
\be 
{\cal M}^2_1 = 2 \, m_\Sigma^2  + \mu_\Sigma \, s_1    ~~.
\ee
The minimization condition ${\cal M}^2_1 > 0 $ is equivalent to
\be
\frac{1}{3 \lambda + \kappa}  \left(  \sqrt{4(3 \lambda + \kappa) \, m^2_\Sigma + \mu^2_\Sigma} \,  \pm  \mu_\Sigma \right)  >  0 ~~,
\label{eq:M1cond}
\ee
where the $+$ or $-$ sign corresponds to the sign chosen for the extremum (\ref{eq:SU3minimum}).
This condition can never be satisfied by the negative solution  (since $m^2_\Sigma  < 0$  in that case), which thus is at most a saddle point.

The minimization condition  (\ref{eq:M1cond}) is automatically satisfied by the positive solution [given the constraint (\ref{eq:3lambda}) in that case], so only ${\cal M}^2_3 > 0 $ remains to be imposed:
\be
(3 \lambda + 2\kappa)  \left(  \sqrt{4(3 \lambda + \kappa) \, m^2_\Sigma + \mu^2_\Sigma} \,  +  \mu_\Sigma \right)  > -2 \kappa (3 \lambda + \kappa) \frac{m^2_\Sigma}{\mu_\Sigma}  ~~.
\ee
For $m^2_\Sigma  > 0$ we find that the positive solution from (\ref{eq:SU3minimum}) represents a local minimum if and only if either 
$\kappa \ge 0$ or else 
\be
\kappa < 0   \;\;\;\;\;   {\rm and} \;\;\;\; \;   \left( 3\lambda + 2 \kappa \right)  \mu_{\Sigma}^2  >  \kappa^2 m_\Sigma^2       ~~.
\ee
For $m^2_\Sigma  \leq 0$ the positive solution is a local minimum when 
\bear
&&   - 3 \lambda   < \kappa    \;\;\;\;\;   {\rm and} \;\;\;\;    \lambda  < 0   ~~~,
\nonumber \\  
&&  \hspace*{-0.5cm}  {\rm  or}
\nonumber \\   [-1mm]
&&  - \frac{3}{2} \lambda < \kappa     \;\;\;\;\;   {\rm and} \;\;\;\;    \lambda  > 0   ~~~,
\nonumber \\  [1mm]
&&  \hspace*{-0.5cm}  {\rm  or}
  \\    [-2mm]
&&  - 2 \lambda < \kappa < -\frac{3}{2} \lambda  < 0        \;\;\;\;\;   {\rm and} \;\;\;\;      \left( 3\lambda + 2 \kappa \right)  \mu_{\Sigma}^2  >  \kappa^2 m_\Sigma^2      ~~~.
\nonumber
\label{eq:SU3cond}
\eear
To derive the above conditions we used the constraints (\ref{eq:mu-constraint}) and (\ref{eq:3lambda}).

The value of the potential at the $SU(3)$-symmetric vacuum is given by  \\ [-1mm]
\be
 V_{\rm min}^{(3)} = -  \frac{3 }{2\left( 3\lambda + \kappa \right)}  \left[ m_\Sigma^4 +  \frac{ m_\Sigma^2 \,  \mu_{\Sigma}^2 }{3\lambda + \kappa} 
+  \frac{ \mu_{\Sigma}^4 +  \mu_{\Sigma} \left(  4(3 \lambda + \kappa) \, m^2_\Sigma + \mu^2_\Sigma   \rule{0pt}{3.7mm} \right)^{\! 3/2} \! \! }{ 6 (3\lambda + \kappa)^2}    \;\;
\right]   ~~.
\label{eq:VA}
\ee
We will discuss the conditions for a global minimum in Section~\ref{sec:global-minimum}. 

Among the 18 degrees of freedom in $\Sigma$, there are 8 exactly massless Nambu-Goldstone Bosons (NGB's). The remaining 10 degrees of freedom are massive and can be decomposed into $8+1+1$ under the unbroken $SU(3)$ vacuum symmetry \cite{Bai:2010dj}. 

\bigskip\bigskip

\section{$SU(2)\times SU(2) \times U(1)$-symmetric vacuum }
\label{sec:su221-vacuum}
\setcounter{equation}{0}

We now seek minima with  two of the $s_i$ vanishing, so that the VEV preserves an $SU(2)\times SU(2) \times U(1)$ symmetry. It is sufficient to set $s_1 > 0$ and $s_2 = s_3 = 0$, as this is equivalent up to  $SU(3)_1 \times SU(3)_2$ transformations to the cases $s_1 = s_2 = 0$ or $s_1 = s_3 = 0$.
Another transformation, along the diagonal generators, can be used in this case to eliminate  the phase $\alpha$ from the VEV (\ref{eq:generalVEV}). 
The extremization conditions (\ref{eq:setDif}) take a simple form,
\be
 \left( \lambda + \kappa \right)   s_1^2  = m_\Sigma^2    ~~~.
\ee
For $( \lambda + \kappa) m_\Sigma^2 > 0$  the extremum is at 
\be
s_1 =  \frac{ |m_\Sigma| }{\sqrt{ |\lambda + \kappa| }}   ~~.
\label{eq:minBs1}
\ee

Using the same rotation on the second derivative matrix  $\partial^2 V/(2\partial s_i\partial s_j)$ 
as in Eq.~(\ref{eq:massSquared-matrix}), we find the eigenvalues 
\bear
& &  {\cal M}^2_1 = 2 m_\Sigma^2   ~~,
\nonumber \\ [2mm]
& &  {\cal M}^2_2 =   - \, \frac{ \kappa }{ \lambda +  \kappa} \,  m_\Sigma^2  -  \mu_\Sigma \,s_1   ~~,
\nonumber \\ [2mm]
& &   {\cal M}^2_{3} =   - \, \frac{ \kappa }{ \lambda +  \kappa} \, m_\Sigma^2  + \mu_\Sigma \, s_1   ~~.
\eear
The minimization condition  ${\cal M}^2_1 > 0$ is satisfied provided  $m_\Sigma^2 > 0$, which implies $ \kappa > - \lambda$.
As $m_\Sigma$ is real and its sign is irrelevant, we choose $m_\Sigma > 0$.
Given that  ${\cal M}^2_{3} >  {\cal M}^2_2$, it remains to impose ${\cal M}^2_2 > 0$, so that
\be
 - \lambda <  \kappa < 0   \;\;\;\;\;   {\rm and} \;\;\;\;  
 \mu_\Sigma  < -  \frac{\kappa \, m_\Sigma}{\sqrt{\lambda + \kappa}} 
  ~~.
  \label{eq:221cond}
\ee
Thus, an $SU(2)\times SU(2) \times U(1)$-symmetric  local minimum exists at 
\be
s_1 =  \frac{ m_\Sigma }{\sqrt{ \lambda + \kappa }}  \;  \; \;  , \;   \;  \;    s_2 = s_3 = 0 ~~.
\label{eq:minB}
\ee
The value of the potential at this minimum is  
\be
V_{\rm min}^{(2,2,1)}       
 = -  \frac{m_\Sigma^4 }{2\left( \lambda + \kappa \right)} < 0  ~~.
\label{eq:SU2minPotential}
\ee

The degrees of freedom in the $\Sigma$ field are grouped into 9 massless NGB's and 9 massive scalars. The latter can be decomposed into a complex scalar 
transforming as $(2, 2, 0)$ under the unbroken $SU(2)\times SU(2)\times U(1)$ vacuum symmetry, and a real singlet scalar.

\smallskip

\section{Absence of  less symmetric vacua}  
\label{sec:less-vacuum}
\setcounter{equation}{0}

Let us now seek extrema with two of the $s_i$ equal but nonzero, so that the remaining symmetry of the $\Sigma$ VEV is the diagonal 
$SU(2) \times U(1) $ subgroup of $SU(3)_1 \times SU(3)_2$. It is sufficient to  consider the case
\be
s_2 = s_3 >  0  \;\; \;  {\rm and}  \;\;\;
s_2 \neq s_1 \geq  0   ~~,
\label{eq:SU2U1} 
\ee
because $SU(3)_1 \times SU(3)_2$ transformations can connect this extremum to 
the ones with permutations of  the $i = 1,2,3$ indices ($s_3 \neq s_1 = s_2 >0  $ or $ s_2 \neq s_1 = s_3 > 0$).
The extremization conditions Eq.~(\ref{eq:setDif}) and (\ref{eq:phase}) are in this case given by  
\bear 
&& \hspace*{-0.81cm}
s_1 \, \sin  \alpha = 0     ~~,
\nonumber \\ [2mm]
&& \hspace*{-0.81cm}
 \left( \lambda \!+\! \kappa \right)  \left( s_1^2 + s_2^2 + s_1 s_2  \right)   +  \lambda \, s_2 \left( s_2 - s_1  \right)  +   \mu_\Sigma \, s_2 \cos  \alpha  = m_\Sigma^2   ~~,
\nonumber \\ [3mm]
&&  \hspace*{-0.81cm}
 \left( \lambda  \!+\!  \kappa \right)  s_2^2  +  \lambda  \left( s_1^2 +  s_2^2 \right)  -  \mu_\Sigma \, s_1 \cos  \alpha = m_\Sigma^2    ~~.
\label{eq:extremeSU2}
\eear

The solution $s_1 = 0$ to the first equation implies $\cos \alpha = 0$, due to the last two equations above. At this extremum, the second-derivative 
matrix [see Eq.~(\ref{eq:secondDerivative})]  is block diagonal, with one of the $2\times 2$ blocks having the determinant equal to $ - s_2^4 < 0$. Thus, at  least one of the eigenvalues is negative
so that 
the extremum at  $s_1 = 0$
is only a saddle point. 

The other solution to the first equation (\ref{eq:extremeSU2}), $\sin \alpha = 0$, leads to more complications. One of the eigenvalues of the second-derivative matrix is given by its 44 entry, 
and is positive only for $\cos\alpha = 1$. 
Imposing this condition as well as  the positivity condition (\ref{eq:SU2U1}), we find that the extremization conditions (\ref{eq:extremeSU2}) have a solution,
\bear
&& s_1 = - \frac{\mu_\Sigma }{\kappa} > 0 ~~,
\nonumber \\ [2mm]
&& s_2= s_3 =  \left(\frac{\kappa^2 m_\Sigma^2 - ( \lambda + \kappa) \mu_\Sigma^2 }{ \kappa^2 (2\lambda + \kappa ) } \right)^{\! 1/2} > 0   ~~,
\nonumber \\ [2mm]
&& \alpha = 0  ~~,
\label{eq:extremumSU2}
\eear
only for 
\be
\kappa < 0    \;\;\;\;  {\rm and} \;\;\;\;   \frac{1}{2\lambda + \kappa}\left[ (\lambda + \kappa )  \mu_\Sigma^2  -  \kappa^2 m_\Sigma^2 \right] < 0   ~~~.
 \label{eq:conditionSU2}
\ee

To see if the extremum  (\ref{eq:extremumSU2})  may be a minimum, we use the mass-squared matrix $ {\cal M}^2$ of Eq.~(\ref{eq:massSquared-matrix}), which in this case 
has  the following nonzero elements:  
\bear
& &  {\cal M}^2_{11} = \frac{  1 }{ 2 \lambda +  \kappa} \left[ - \kappa  \, m_\Sigma^2  +  (\lambda + \kappa)  (4\lambda + 3\kappa) \frac{  \mu_\Sigma^2}{  \kappa^2}   \right]  ~~,
\nonumber \\ [3mm]
& &  {\cal M}^2_{22} =   2m_\Sigma^2  -  2  (\lambda + \kappa)  \frac{ \displaystyle \mu_\Sigma^2}{ \displaystyle \kappa^2}    ~~,
\nonumber \\ [3mm]
&&   {\cal M}^2_{12} =   -\sqrt{2} (2 \lambda +\kappa )  \frac{  \displaystyle \mu_\Sigma}{ \displaystyle \kappa} s_2     ~~,
\nonumber \\ [3mm]
& &   {\cal M}^2_{3} =    \frac{ 2\kappa }{ 2 \lambda +  \kappa} \left[  m_\Sigma^2  - (3\lambda + 2\kappa) \frac{ \mu_\Sigma^2}{  \kappa^2}   \right]      ~~.
\eear
The determinant of $ {\cal M}^2$ is given by $ - (2 \lambda +  \kappa) s_2^2 {\cal M}^4_{3}$, so a necessary minimization condition 
is 
\be 
2 \lambda +  \kappa < 0  ~~,
\ee
which in conjunction with  (\ref{eq:conditionSU2})  implies  
$\lambda + \kappa < 0 $ and  $m_\Sigma^2 < 0$. 
Another necessary minimization condition is $ {\cal M}^2_{11}  +  {\cal M}^2_{22} > 0 $, which leads to 
\be
(\lambda + \kappa )  \mu_\Sigma^2  >  - \kappa (4 \lambda + \kappa )   m_\Sigma^2  ~~.
\label{eq:strongerCond}
\ee
The remaining minimization condition is $  {\cal M}^2_{3} > 0 $, implying
\be
(3\lambda + 2\kappa )  \mu_\Sigma^2  <  \kappa^2 m_\Sigma^2  ~~,
\label{eq:M3}
\ee
which is incompatible with (\ref{eq:strongerCond}).  Thus, the solution  (\ref{eq:extremumSU2})  is  only a  saddle point. 

Let us finally seek solutions to the  extremization conditions (\ref{eq:setDif}) and  (\ref{eq:phase})   where  $s_i \neq s_j $ for all $i\neq j$, with $i, j = 1,2,3$.
Note that when $s_2 \neq s_3$ the last two equations in (\ref{eq:setDif})  are equivalent to
\bear
&& \kappa \, s_2 s_3  = - \mu_\Sigma \, s_1 \cos\alpha  ~~~,
\nonumber \\ [2mm]
&&  \left( \lambda + \kappa \right)  \left( s_2^2 + s_3^2 \right)  = m_\Sigma^2 -  \lambda  \, s_1^2   ~~.
\label{eq:asymmetricEqs}
\eear
From Eq.~(\ref{eq:first-deriv}) it follows
\be
s_1^2 = \frac{1 }{ 2 \lambda + \kappa }  \left( m_\Sigma^2 - (\lambda + \kappa)  \frac{   \mu_{\Sigma}^2}{  \kappa^2 }   \cos^2\!\alpha   \right) ~~.
\label{eq:asymmetric1}
\ee
As at most one $s_i$ vanishes,  we can take $s_1, s_2 > 0$, so Eq.~(\ref{eq:phase}) becomes  $ \mu_\Sigma \,  s_3 \, \sin  \alpha  = 0 $.
The solution with $s_3 = \cos\alpha = 0$ is not allowed because Eqs.~(\ref{eq:asymmetricEqs}) and (\ref{eq:asymmetric1})
 imply $s_1 = s_2$. 
The solution  with $s_3 > 0$ and  $ \sin  \alpha = 0$ is less obvious, but it also leads to $s_1 = s_2$. 
Thus, there is no extremum when all three $s_i$ are different.

\bigskip

\section{Asymptotic behavior}
\label{sec:asymptotic}
\setcounter{equation}{0}

A necessary condition for the existence of a global minimum is that there are no runaway directions at large field values. 
In other words, $V(\Sigma)$ must have a lower limit as $s_i \to \infty$. 

At large field values, where  the $\mu_\Sigma$ and $m_\Sigma$ terms  can be neglected, the potential (\ref{eq:sigma-pot}) has  the following asymptotic form:
\be
V_\infty = \frac{\lambda}{2}   \left(s_1^2 + s_2^2 + s_3^2\right)^2 
\,+\, \frac{\kappa}{2} \left(s_1^4 + s_2^4 + s_3^4\right) ~~\,.
\ee
Hence,  in the case where  $s_1 = s_2 = s_3 \to  \infty$,  the condition that 
$V(\Sigma)$ is bounded from below is $\, 3\lambda + \kappa > 0$ (this was also derived in \cite{Chivukula:2013xka}).
We point out that a separate condition for $V(\Sigma)$ to be bounded from below is obtained 
in the case where  $s_i \to  \infty$ for a single value of $i$:  
$\, \lambda + \kappa > 0$.   
These two conditions can be combined as follows: 
\be
 \kappa >   {\rm max}  \left\{ - \lambda , - 3 \lambda \right\}  ~~,
 \label{eq:bounded}
\ee
which is a  necessary condition to have $V(\Sigma)$ bounded from below. 

We now prove that (\ref{eq:bounded}) is also a sufficient condition to have a bounded potential. For $\lambda \ge 0$, the condition becomes $\kappa > - \lambda$ so that 
\be
V_\infty  > \frac{\lambda}{2}   \left(s_1^2 + s_2^2 + s_3^2\right)^2 
\,-\, \frac{\lambda}{2} \left(s_1^4 + s_2^4 + s_3^4\right) 
= \lambda \left( s_1^2 s_2^2 + s_1^2 s_3^2 + s_2^2 s_3^2 \right)  \ge 0 ~~\,.
\ee
For $\lambda < 0$, condition (\ref{eq:bounded}) becomes $\kappa > - 3 \lambda > 0$, 
which implies
\bear
V_\infty & > &  - \frac{\kappa}{6} \, \left(s_1^2 + s_2^2 + s_3^2\right)^2
 \,+\,\frac{\kappa}{2}  \left(s_1^4 + s_2^4 + s_3^4\right) 
 \nonumber \\ [2mm]
&=& \frac{\kappa}{12} \left[ \left( 2 s_1^2 - s_2^2 - s_3^2 \right)^2 + 3\left( s_2^2 - s_3^2 \right)^2 \right] \ge 0   ~~.
\eear
Therefore, (\ref{eq:bounded}) is the sufficient and necessary condition to have $V(\Sigma)$ bounded from below.  

\bigskip

\section{Global minimum}
\label{sec:global-minimum}
\setcounter{equation}{0}

As established in Sections 2-5, the renormalizable potential for a single bifundamental scalar allows only three possible vacua:
\bear
&& \hspace*{-1.cm} 
SU(3)\times SU(3) \;\; {\rm vacuum: }  \;\;\;\;  s_1 = s_2 = s_3 = 0
 \nonumber \\ [2mm]
&& \hspace*{-1.cm}
SU(3) \;\; {\rm vacuum: }  \;\;\;\;  s_1 =  s_2 = s_3 =  
\frac{1}{ 2 (3 \lambda + \kappa)} \left( \sqrt{4(3 \lambda + \kappa) \, m^2_\Sigma + \mu^2_\Sigma} \, + \mu_\Sigma \right)   \;   ,  \;  \;   \alpha = 0
 \nonumber \\ [2mm]
&& \hspace*{-1.cm}
SU(2)\times SU(2) \times U(1) \;\;  {\rm vacuum: }  \;\;  s_1 =  \frac{ m_\Sigma }{\sqrt{ \lambda + \kappa }}  \;  \;  ,  \;  \;    s_2 = s_3 = 0 ~.
\eear
Let us analyze which of these local minima represents a 
global minimum of the potential. To this end we need to impose first the condition that $V(\Sigma)$ is bounded from below, namely (\ref{eq:bounded}).
In this case the regions of parameter space where the $SU(3)$-symmetric  and  $SU(2)\times SU(2) \times U(1) $-symmetric vacua exist, namely 
  (\ref{eq:SU3cond}) and (\ref{eq:221cond}),  are simpler. 
Three regions of parameter space  have a single vacuum: 
\bear
&& \hspace*{-1.cm}
m_\Sigma > 0  \; , \;    \kappa < 0   \;\;\;  {\rm and} \;\;\;
 \frac{ \mu_\Sigma }{m_\Sigma}  <   \frac{ - \kappa }{\sqrt{ 3\lambda + 2\kappa}}   \;\;\;   \Rightarrow  \;\;\;    SU(2)\times SU(2) \times U(1) \;\; {\rm vacuum} 
 \nonumber \\ [3mm]
&&  \hspace*{-1.cm}
m_\Sigma > 0  \;\;\;  {\rm and} \;\;\;   \Big\{ \kappa \geq 0   \;\;  {\rm or}  \;\;   \frac{\mu_\Sigma}{m_\Sigma}  > \frac{- \kappa}{\sqrt{\lambda + \kappa}}  \Big\}  
 \;\;\;   \Rightarrow  \;\;\;  SU(3)   \;\; {\rm vacuum}  
 \nonumber \\ [3mm]
&&  \hspace*{-1.cm}
m_\Sigma^2 \! < 0   \;\;\;  {\rm and} \;\;\;  
 \frac{- \mu_\Sigma^2}{m_\Sigma^2 \left( 3\lambda \!+\! \kappa \right) \! }  < 4 
 \;\;\;   \Rightarrow  \;\;\;   SU(3) \times SU(3)   \;\; {\rm vacuum}  
  \label{eq:simpleVac}
\eear 
where again we chose $m_\Sigma >0$ when $m_\Sigma^2 > 0$. 

In the other regions there is competitions between two vacua.  Studying the sign of the potential at the $SU(3)$-symmetric minimum, $V_{\rm min}^{(3)}$ of Eq.~(\ref{eq:VA}),
we find\footnote{This result agrees with the one derived in Appendix A of  Ref.~\cite{Chivukula:2013xka}, namely $SU(3)$ global minimum at  $r_\Delta < 3/2$ in the notation used there. 
The competition between the $SU(2)\times SU(2) \times U(1)$-symmetric  minimum and the $SU(3)$-symmetric minimum is not discussed in Ref.~\cite{Chivukula:2013xka}.}
\bear
&&  \hspace*{-0.6cm}
m_\Sigma^2 \! < 0   \; \; {\rm and} \; \;  \frac{9}{2} < \frac{- \mu_\Sigma^2}{ m_\Sigma^2 \left( 3\lambda \!+\! \kappa \right) \! } 
 \;  \Rightarrow  \;
 \left\{ \ba{l} SU(3)    \;\; {\rm global \; min.}   \\ [4mm] 
\! SU(3)\!\times\! SU(3)    \;\; {\rm local \; min.}    \ea \right. 
 \nonumber  \\ [4mm]
  && \hspace*{-0.6cm}
m_\Sigma^2 \! < 0   \; \; {\rm and} \; \; 
4 <  \frac{- \mu_\Sigma^2}{ m_\Sigma^2 \left( 3\lambda \!+\! \kappa \right) \! }  < \frac{9}{2} 
 \;  \Rightarrow  \;
 \left\{ \ba{l} SU(3)\!\times\! SU(3)    \;\; {\rm global \; min.}   \\ [4mm] 
\! SU(3)    \;\; {\rm local \; min.}    \ea \right. 
 \nonumber  \\ [-6mm]
\label{eq:3vs33}
\eear 

For the remaining region of parameter space, 
\be
  \hspace*{-0.51cm}
m_\Sigma > 0  \; , \;    \kappa < 0     \;\;\;   {\rm and} \;\;\;  
\frac{ - \kappa }{\sqrt{ 3\lambda + 2\kappa}}  <   \frac{\mu_\Sigma}{m_\Sigma}  <   \frac{-\kappa }{\sqrt{\lambda + \kappa}}  ~~,
\ee
 there is competition between the  $SU(3)$ and $ SU(2)\times SU(2) \times U(1)$  local minima.
We need to compare the values of the potential at these minima, which are given in Eqs.~(\ref{eq:VA}) and (\ref{eq:SU2minPotential}). 
 The $SU(3)$ minimum  is deeper,  $V_{\rm min}^{(3)}  <   V_{\rm min}^{(2,2,1)}  $,
if and only if \\ [-1mm]
\be 
\frac{\mu_\Sigma}{m_\Sigma}   \, > \, \left(  \frac{ \left( 4\lambda + 2\kappa \right)^{3/2} \!\! \!}{ \sqrt{\lambda + \kappa}} \, -2(4\lambda +  \kappa)   \right)^{\!\! 1/2}
 \equiv  \xi(\lambda,\kappa)  
  ~~.
\label{eq:xi}
\ee
One can check that the function defined above, $\xi(\lambda,\kappa)$, is real and positive in this region of parameter space. As a result, we find the following possible vacua:
\bear
&&  \hspace*{-0.6cm}
m_\Sigma \!  > 0  \, , \,  \kappa \! < 0     \;\;   {\rm and} \;\;
\xi(\lambda,\kappa) <  \frac{ \mu_\Sigma }{m_\Sigma}  <   \frac{-\kappa}{\sqrt{\lambda + \kappa}}  
 \;  \Rightarrow  \;
 \left\{ \ba{l} SU(3)    \;\; {\rm global \; min.}   \\ [4mm] 
\! SU(2)\!\times\! SU(2) \!\times\! U(1)    \;\; {\rm local \; min.}    \ea \right. 
 \nonumber  \\ [5mm]
&&  \hspace*{-0.6cm}
m_\Sigma \!  > 0  \, , \,   \kappa \! < 0     \;\;   {\rm and} \;\;
 \frac{-\kappa}{\! \sqrt{3\lambda + 2\kappa}}  < \frac{ \mu_\Sigma }{m_\Sigma} < \xi(\lambda,\kappa) 
 \;  \Rightarrow 
 \left\{ \ba{l} \! \! SU(2)\!\times\! SU(2) \!\times\! U(1)    \; {\rm global \; min.}   \\ [4mm] 
SU(3)    \;\; {\rm local \; min.}    \ea \right. 
 \nonumber \\  [-4mm]
 && \hspace*{-8.cm}
 \label{eq:3vs221}
\eear 

\begin{figure}[t]
\begin{center}
\includegraphics[width=0.68\textwidth]{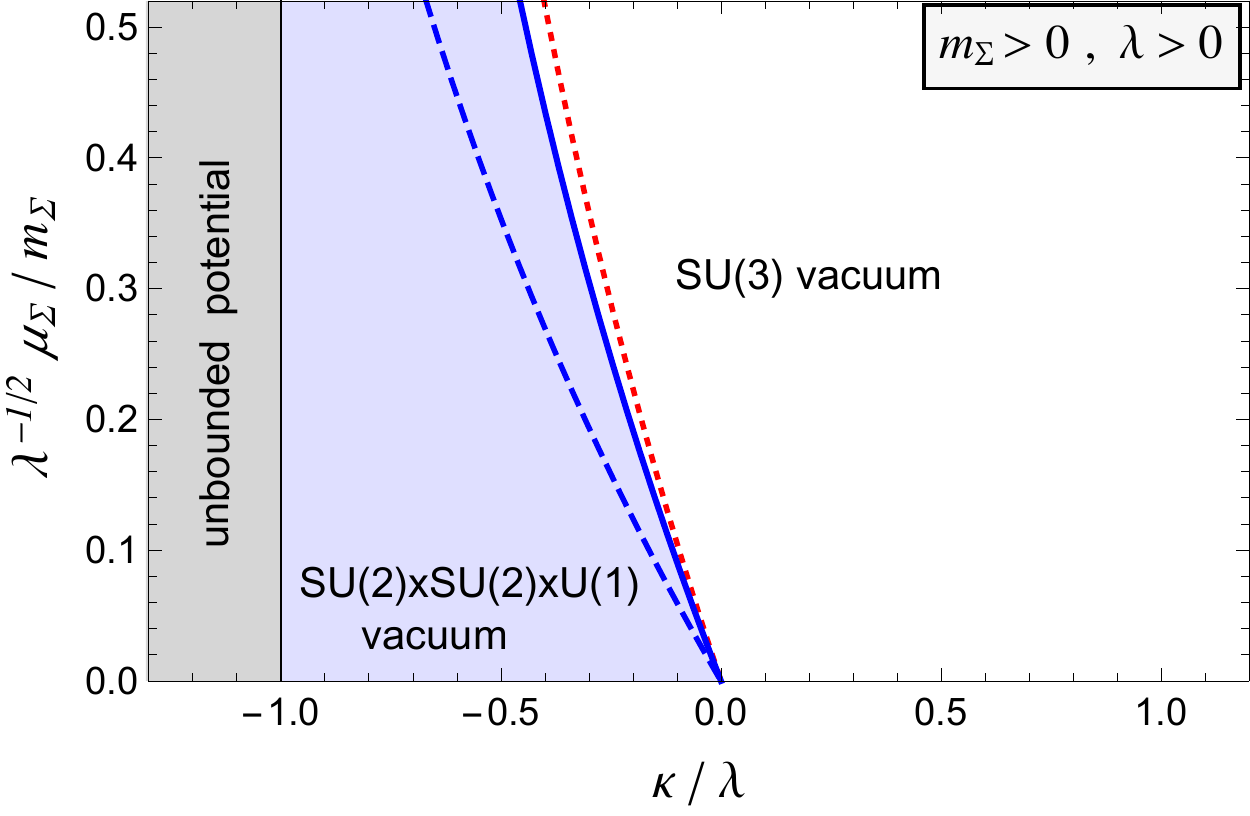} 
\caption{Phase diagram of the $SU(3)\times SU(3)$ model 
with a scalar bifundamental, for $m_\Sigma^2 > 0$ and $\lambda > 0$, 
in the plane of $\mu_\Sigma / (m_\Sigma \sqrt{\lambda}) $ versus the ratio of quartic couplings $\kappa/\lambda$.
The global minimum is $SU(2)\times SU(2) \times U(1) $-symmetric 
in the blue shaded region, and $SU(3)$-symmetric in the unshaded region. Between the dashed blue  line and the solid blue line there is also an $SU(3)$-symmetric local minimum,
while between the dotted red line and the solid blue line there is also an $SU(2)\times SU(2) \times U(1)$-symmetric local minimum.
In the gray-shaded  region at $\kappa/\lambda < -1$ the potential is not bounded from below.
 }
\label{fig:phase}
\end{center}
\end{figure}
\begin{figure}[t]
\begin{center}
\includegraphics[width=0.68\textwidth]{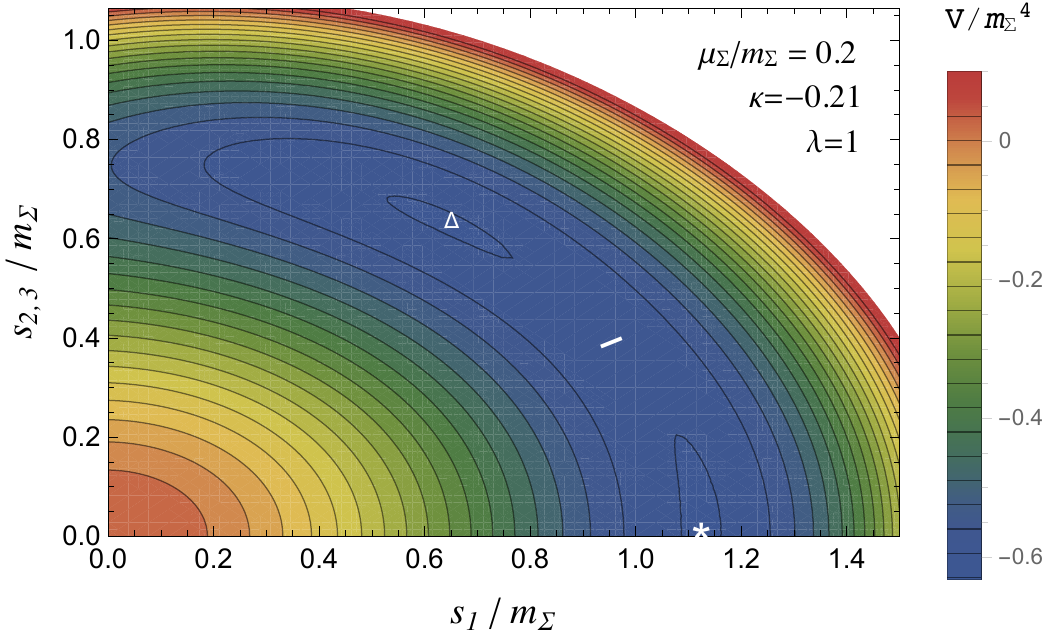} 
\caption{Contours of the $V(\Sigma)$ potential in the ($s_1/m_\Sigma \, , s_2/m_\Sigma$) plane, along the $s_2 = s_3$ and $\alpha = 0$ direction. 
The depth of the potential is encoded in the colors: from  dark blue representing deepest, to bright red representing highest. 
The potential is computed  at a point in parameter space  ($\mu_\Sigma / m_\Sigma = 0.2$, $\kappa = -0.21$, $\lambda = 1$) chosen such that 
the $SU(3)$-symmetric minimum (marked by a white {\small $\triangle$}) 
and the $SU(2)\times SU(2) \times U(1)$-symmetric minimum  (marked by a white {\large $\ast$}) 
have equal depths, of $-0.632 \, m_\Sigma^4$. The saddle point between the two minima  (marked by a white tilted line) has an $SU(2) \times U(1)$
symmetry and a depth of $-0.627 \, m_\Sigma^4$. 
 }
\label{fig:phase2}
\end{center}
\end{figure}

\smallskip

The phase diagram of this model, based on Eqs.~(\ref{eq:simpleVac}),  (\ref{eq:3vs33})  and (\ref{eq:3vs221}),
 is shown in Figure~\ref{fig:phase} in the $\lambda^{-1/2} \mu_\Sigma / m_\Sigma$ versus $\kappa/\lambda$ plane, for $m_\Sigma > 0$ and $\lambda >0$.
Note that for $\lambda >0$ the lower limit
$\kappa/\lambda > -1$ is required in order to have the potential bounded from below, while there is no upper limit on $\kappa/\lambda$ at tree level.

The region where the global minimum is $SU(2)\times SU(2) \times U(1)$-symmetric lies below the solid blue line in Figure~\ref{fig:phase}, which is given by 
the function  $\xi(\lambda,\kappa)/\sqrt{\lambda}$    [see Eq.~(\ref{eq:xi})]. 
In the region above or to the right of that line, the global minimum is $SU(3)$-symmetric.

\bigskip\bigskip

A change of parameters that crosses the boundary between these two regions represents  a first-order phase transition: both local minima exist 
for parameter points between the blue dashed line and the red dotted line of Figure~\ref{fig:phase}. In between these two minima there is a shallow 
saddle point, of coordinates given in (\ref{eq:extremumSU2}), which is $SU(2)\times U(1)$-symmetric. 
In Figure~\ref{fig:phase2} we show the potential for a point ($\mu_\Sigma / m_\Sigma = 0.2$, $\kappa = -0.21$, $\lambda = 1$) from the  $\xi(\lambda,\kappa)/\sqrt{\lambda}$ curve,
where the $SU(3)$-symmetric vacuum and the $SU(2)\times SU(2) \times U(1)$-symmetric vacuum have the same depth and are global minima. The shallowness of the potential around both minima is related to the smallness of $|\mu_\Sigma/m_\Sigma|$ and $|\kappa/\lambda|$. The mass of the ``angular mode" is parametrically smaller than the ``radial mode".

\begin{figure}[t]
\begin{center}
\includegraphics[width=0.68\textwidth]{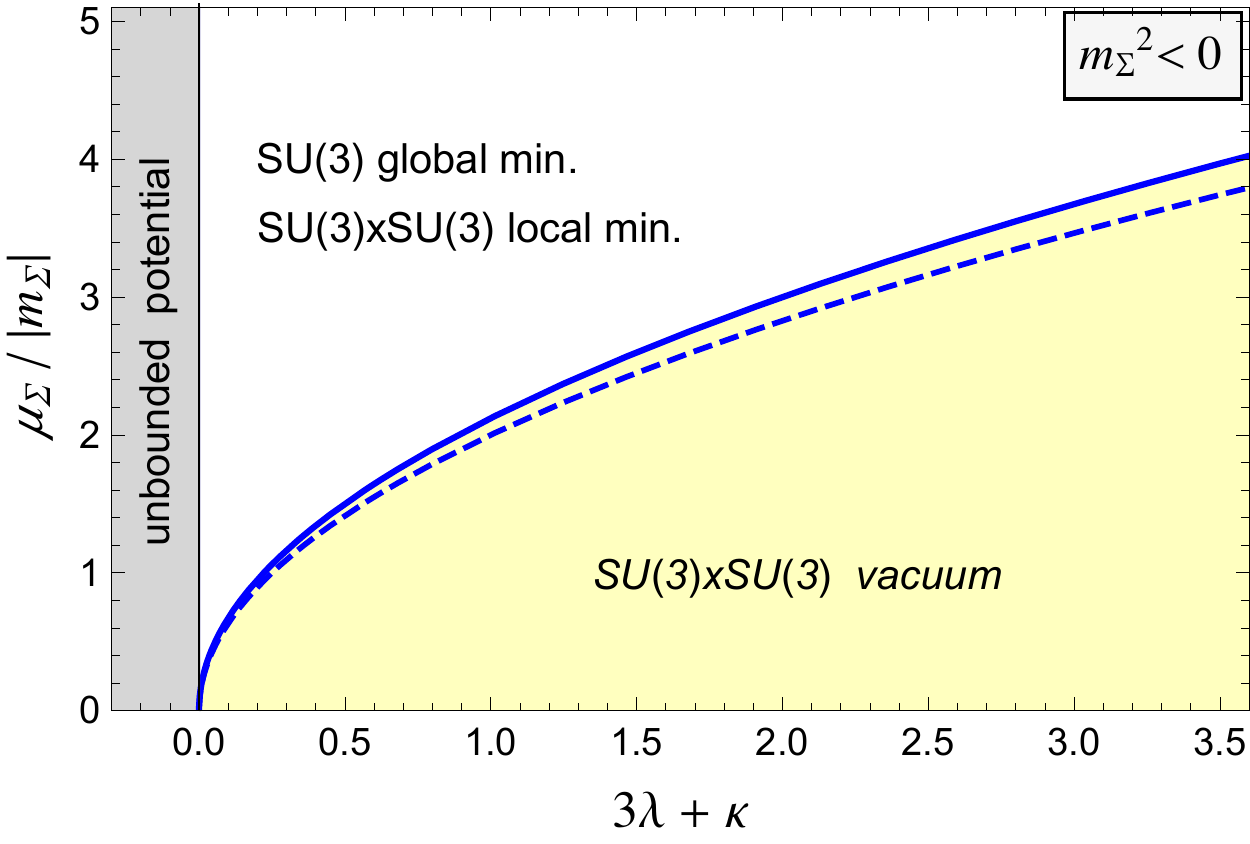} 
\caption{Phase diagram for $m_\Sigma^2 < 0$, in the plane of $\mu_\Sigma / |m_\Sigma| $ versus $3 \lambda + \kappa$.
The global minimum is $SU(3)$-symmetric in the unshaded region, and 
$SU(3)\times SU(3)$-symmetric in the yellow shaded region.
Between the dashed blue  line and the solid blue line there is also an $SU(3)$-symmetric local minimum,
while in the unshaded region there is also an $SU(3)\times SU(3)$-symmetric local minimum.
In the gray-shaded  region at $3 \lambda + \kappa < 0$ the potential is not bounded from below.
 }
\label{fig:phase3}
\end{center}
\end{figure}

The region where $m_\Sigma > 0$ and $\lambda < 0$ has only the $SU(3)$-symmetric vacuum. In the phase diagram for $m_\Sigma^2 < 0$,
shown in Figure~\ref{fig:phase3}, there is competition between the $SU(3)\times SU(3)$ vacuum and the $SU(3)$ vacuum, 
as described by the inequalities (\ref{eq:simpleVac}) and  (\ref{eq:3vs33}). On the boundary between the two regions defined in (\ref{eq:3vs33}), given by the solid blue line in Figure~\ref{fig:phase3}, the two minima are degenerate. The saddle point that separates these two global minima 
corresponds to the negative-sign solution of Eq.~(\ref{eq:SU3minimum}). 

\begin{figure}[t]
\begin{center}
\includegraphics[width=0.68\textwidth]{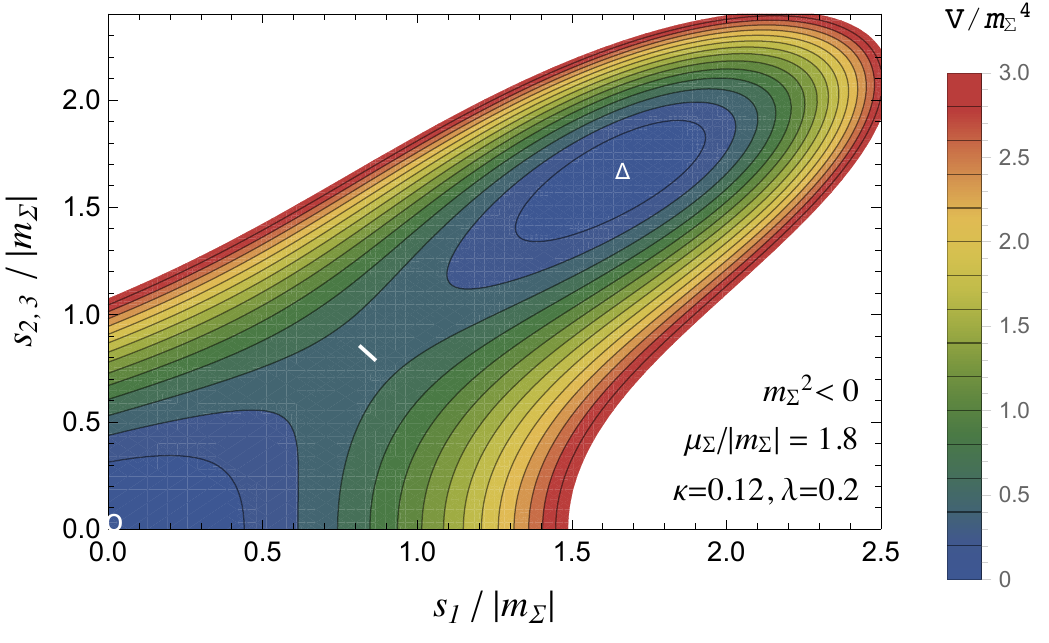} 
\caption{Same as Figure 2, except that the potential is computed  at a point in parameter space ($m_\Sigma^2<0$ and 
$\mu_\Sigma / |m_\Sigma| = 1.8$, $\kappa = 0.12$, $\lambda = 0.2$) chosen such that 
the $SU(3)$-symmetric minimum (marked by a white {\small $\triangle$}) 
and the $SU(3)\times SU(3)$-symmetric minimum  (marked by a white {\large $\circ$}) 
have equal depths. 
The saddle point between the two minima  (marked by a white tilted line) has the same $SU(3)$
symmetry, and corresponds to the negative-sign solution of Eq.~(\ref{eq:SU3minimum}). }
\label{fig:phase4}
\end{center}
\end{figure}

In Figure~\ref{fig:phase4}, we show the potential for a point with $m_\Sigma^2<0$, located on the boundary
at $\mu_\Sigma / |m_\Sigma| = 1.8$, $\kappa = 0.12$, $\lambda = 0.2$, where the depth of the potential is same at the two minima ($V=0$), and at the saddle point it is given by 
$V= 0.52 \, m_\Sigma^4$.


Note that the inequalities (\ref{eq:simpleVac}),  (\ref{eq:3vs33})  and (\ref{eq:3vs221}) do not explicitly refer to
the cases where some parameters vanish. The reason for that is that the analysis 
in those cases becomes sensitive to loop corrections.\footnote{The computation of the 1-loop effective potential is a 
mature subject (see, {\it e.g.}, \cite{Camargo-Molina:2013qva}), and even the 3-loop  
effective potential has been recently computed for a general renormalizable theory \cite{Martin:2017lqn}.}
For example, $\lambda = 0$ at tree level makes the vertical axis 
ill defined in Figure~\ref{fig:phase}, but 1-loop corrections would generate a nonzero $\lambda$. Likewise, $m_\Sigma = 0$ is not 
stable against loops. By contrast, the $\mu_\Sigma = 0$ limit is protected by a global $U(1)_\Sigma$ symmetry, as discussed in Section \ref{sec:potential}.

We emphasize that although the global minimum of the potential will eventually be the vacuum, 
the universe might be stuck for a while in the shallower local minimum.  Thus, a local minimum may be a viable vacuum provided that it is 
longer-lived than the age of the universe, and that the thermal history allows the universe to settle in it.

\vspace*{0.1cm}

\vfil
\newpage\newpage

\section{Conclusions} 
\label{sec:coclusions}
\setcounter{equation}{0}

We have analyzed the vacuum structure of an $SU(3)\times SU(3)$-symmetric renormalizable theory with a bifundamental scalar field.
The parameter space is 4-dimensional, with two quartic couplings and two mass parameters. One of the latter, which is the coefficient of a cubic 
term in the potential, is not present in $SU(N)\times SU(N)$-symmetric theories for $N \neq 3$. 

There are three possible types of vacua, with different symmetry properties: $SU(3)$, $SU(2)\times SU(2) \times U(1)$ and $SU(3)\times SU(3)$.
Depending on which of these is a global minimum, and whether there are also some local minima, the parameter space is divided in 7 regions. 
These are described by Eqs.~(\ref{eq:simpleVac}),  (\ref{eq:3vs33})  and (\ref{eq:3vs221}).
Remarkably, the phase diagram of the theory can be fully displayed in two-dimensional plots, namely Figures~\ref{fig:phase} and \ref{fig:phase3}.

The cubic term in the potential, of coefficient $\mu_\Sigma > 0$, plays an important role in the selection of the possible vacua. 
Even when the bifundamental scalar has a positive squared-mass, 
{\it i.e.}, $m_\Sigma^2 < 0$ in the notation of Eq.~(\ref{eq:sigma-pot}),
a nontrivial VEV is developed for $\mu_\Sigma$ above a coupling-dependent value (see Figure~\ref{fig:phase3}), breaking the symmetry down to $SU(3)$.
For a negative squared-mass (or equivalently  $m_\Sigma > 0$),  as $\mu_\Sigma$ increases, the region with an $SU(3)$-symmetric vacuum is enlarged, while the region with an $SU(2)\times SU(2) \times U(1)$-symmetric vacuum is reduced (see Figure \ref{fig:phase}).

The vacuum structure of this theory is useful for various model building applications, including in the contexts of the ReCoM \cite{Bai:2010dj, Chivukula:2013xka} and Trinification \cite{Glashow:1984gc, Achiman:1978rv, Babu:1985gi}, or 
 chiral symmetry breaking in strongly-coupled gauge theories \cite{Vafa:1983tf}.
In addition, it opens new possibilities for nonstandard cosmology, such as color-breaking in the early universe followed by color restoration at a lower temperature \cite{Ramsey-Musolf:2017tgh}.  In particular, the presence of two minima of different symmetry properties, which for a range of parameters are nearly degenerate 
and separated by a shallow saddle point  (see Figure~\ref{fig:phase2})  may lead to exotic cosmological or astrophysical phenomena.

\bigskip\bigskip\bigskip\bigskip

{\it  Acknowledgments:}  {\small We thank Kiel Howe, Zackaria Chacko and Daniel Chung for useful discussions and comments. The work of YB is supported by the U. S. Department of Energy under the contract DE-FG-02-95ER40896. The work of BD has been supported by Fermi Research Alliance, LLC under Contract No. DE-AC02-07CH11359 with the U.S. Department of Energy, Office of Science, Office of High Energy Physics. 
}


\vfil

\begin{thebibliography}{99} \frenchspacing

\bibitem{Preskill:1980mz} 
  J.~Preskill,
  ``Subgroup Alignment in Hypercolor Theories,''
  Nucl.\ Phys.\ B {\bf 177}, 21 (1981).

\bibitem{Hall:1985wz} 
  L.~J.~Hall and A.~E.~Nelson,
  ``Heavy Gluons and Monojets,''
  Phys.\ Lett.\  {\bf 153B}, 430 (1985).

\bibitem{Frampton:1987dn} 
  P.~H.~Frampton and S.~L.~Glashow,
  ``Chiral Color: An Alternative to the Standard Model,''
  Phys.\ Lett.\ B {\bf 190}, 157 (1987). \\
  J.~Bagger, C.~Schmidt and S.~King,
  ``Axigluon Production in Hadronic Collisions,''
  Phys.\ Rev.\ D {\bf 37}, 1188 (1988). 

\bibitem{Buschmann:2017ucg} 
  M.~Buschmann and F.~Yu,
  ``Collider constraints and new tests of color octet vectors,''
  JHEP {\bf 1709}, 101 (2017)
  [arXiv:1706.07057 [hep-ph]], and reference therein.

  
\bibitem{Hill:1991at} 
  C.~T.~Hill,
  ``Topcolor: Top quark condensation in a gauge extension of the standard model,''
  Phys.\ Lett.\ B {\bf 266}, 419 (1991).

\bibitem{Chivukula:1996yr} 
  R.~S.~Chivukula, A.~G.~Cohen and E.~H.~Simmons,
  ``New strong interactions at the Tevatron?,''
  Phys.\ Lett.\ B {\bf 380}, 92 (1996)
  [hep-ph/9603311]. \\
  E.~H.~Simmons,
  ``Coloron phenomenology,''
  Phys.\ Rev.\ D {\bf 55}, 1678 (1997)
  [hep-ph/9608269].
  
\bibitem{Dobrescu:1997nm} 
  B.~A.~Dobrescu and C.~T.~Hill,
  ``Electroweak symmetry breaking via top condensation seesaw,''
  Phys.\ Rev.\ Lett.\  {\bf 81}, 2634 (1998)
  [hep-ph/9712319]. \\
  R.~S.~Chivukula, B.~A.~Dobrescu, H.~Georgi and C.~T.~Hill,
  Phys.\ Rev.\ D {\bf 59}, 075003 (1999)
  [hep-ph/9809470].

\bibitem{Bai:2010dj} 
  Y.~Bai and B.~A.~Dobrescu,
  ``Heavy octets and Tevatron signals with three or four b jets,''
  JHEP {\bf 1107}, 100 (2011)
  [arXiv:1012.5814 [hep-ph]]. 


\bibitem{Chivukula:2013xka} 
  R.~S.~Chivukula, A.~Farzinnia, J.~Ren and E.~H.~Simmons,
  ``Constraints on the Scalar Sector of the Renormalizable Coloron Model,''
  Phys.\ Rev.\ D {\bf 88}, no. 7, 075020 (2013)
  Erratum: [Phys.\ Rev.\ D {\bf 89}, no. 5, 059905 (2014)]
  [arXiv:1307.1064 [hep-ph]].

  \bibitem{Chivukula:2014rka} 
  R.~S.~Chivukula, E.~H.~Simmons, A.~Farzinnia and J.~Ren,
  ``LHC Constraints on a Higgs boson Partner from an Extended Color Sector,''
  Phys.\ Rev.\ D {\bf 90}, no. 1, 015013 (2014)
  [arXiv:1404.6590 [hep-ph]]. \\
  R.~S.~Chivukula, A.~Farzinnia and E.~H.~Simmons,
  ``Vacuum Stability and Triviality Analyses of the Renormalizable Coloron Model,''
  Phys.\ Rev.\ D {\bf 92}, no. 5, 055002 (2015)
  [arXiv:1504.03012 [hep-ph]]. \\
  R.~S.~Chivukula, A.~Farzinnia, K.~Mohan and E.~H.~Simmons,
  ``Diphoton Resonances in the Renormalizable Coloron Model,''
  Phys.\ Rev.\ D {\bf 94}, no. 3, 035018 (2016)
  [arXiv:1604.02157 [hep-ph]].
        

\bibitem{Kiel}
P. Agrawal and  K. Howe, preprint Fermilab-PUB-17-403-T, October 2017.

\bibitem{Martin:1992aq} 
  S.~P.~Martin,
  ``A Tumbling top quark condensate model,''
  Phys.\ Rev.\ D {\bf 46}, 2197 (1992)
  [hep-ph/9204204].

\bibitem{Cheng:2001nh} 
  H.~C.~Cheng, C.~T.~Hill and J.~Wang,
  ``Dynamical electroweak breaking and latticized extra dimensions,''
  Phys.\ Rev.\ D {\bf 64}, 095003 (2001)
  [hep-ph/0105323].

\bibitem{Pisarski:1983ms} 
  See, {\it e.g.}, R.~D.~Pisarski and F.~Wilczek,
  ``Remarks on the Chiral Phase Transition in Chromodynamics,''
  Phys.\ Rev.\ D {\bf 29}, 338 (1984).

\bibitem{Glashow:1984gc} 
  S.~L.~Glashow,  ``Trinification of all elementary particle forces,''  Print-84-0577 (Boston University), 
in Fifth Workshop on Grand Unification: proceedings, edited by K.~Kang, H.~Fried and P.~Frampton (World Scientific, 1984), p. 88. 

\bibitem{Achiman:1978rv} 
  Y.~Achiman and B.~Stech, in Advanced Summer Institute on New Phenomena in Lepton and Hadron Physics, eds. D. E. C. Fries and J. Wess (Plenum, 1979), p. 303.
  V.~A.~Rizov,
  ``A Gauge Model of the Electroweak and Strong Interactions Based on the Group $SU(3)_L \times SU(3)_R \times SU(3)_c$,''
  Bulg.\ J.\ Phys.\  {\bf 8}, 461 (1981).

\bibitem{Babu:1985gi} 
  K.~S.~Babu, X.~G.~He and S.~Pakvasa,
  ``Neutrino masses and proton decay modes in $SU(3)\times SU(3)\times SU(3)$ Trinification,''
  Phys.\ Rev.\ D {\bf 33}, 763 (1986). \\
  X.~G.~He and S.~Pakvasa,
  ``Baryon Asymmetry in $SU(3)^3 \times Z(3)$ Trinification Model,''
  Phys.\ Lett.\ B {\bf 173}, 159 (1986). \\
  A.~E.~Nelson and M.~J.~Strassler,
  ``Suppressing flavor anarchy,''
  JHEP {\bf 0009}, 030 (2000)
  [hep-ph/0006251].  \\
  S.~Willenbrock,
  ``Triplicated trinification,''
  Phys.\ Lett.\ B{\bf 561}, 130 (2003)
  [hep-ph/0302168].   \\
  K.~S.~Babu, E.~Ma and S.~Willenbrock,
  ``Quark lepton quartification,''
  Phys.\ Rev.\ D {\bf 69}, 051301 (2004)
  [hep-ph/0307380]. \\
  J.~E.~Kim,
  ``Trinification with $\sin^2 \theta_W= 3/8$ and seesaw neutrino mass,''
  Phys.\ Lett.\ B {\bf 591}, 119 (2004)
  [hep-ph/0403196]. \\
  C.~D.~Carone and J.~M.~Conroy,
  ``Higgsless GUT breaking and trinification,''
  Phys.\ Rev.\ D {\bf 70}, 075013 (2004)
  [hep-ph/0407116]; \
  ``Five-dimensional trinification improved,''
  Phys.\ Lett.\ B {\bf 626}, 195 (2005)
  [hep-ph/0507292].
  \\
  A.~Demaria, R.R.~Volkas,
  ``Kink-induced symmetry breaking patterns in brane-world $SU(3)^3$ trinification models,''
  Phys.\ Rev.\ D {\bf 71}, 105011 (2005)
  [hep-ph/0503224].  \\
  J.~Sayre, S.~Wiesenfeldt and S.~Willenbrock,
  ``Minimal trinification,''
  Phys.\ Rev.\ D {\bf 73}, 035013 (2006)
  [hep-ph/0601040]. \\
  B.~Stech,
  ``Trinification Phenomenology and the structure of Higgs Bosons,''
  JHEP {\bf 1408}, 139 (2014)
  [arXiv:1403.2714 [hep-ph]].  \\
  J.~Hetzel and B.~Stech,
  ``Low-energy phenomenology of trinification: an effective left-right-symmetric model,''
  Phys.\ Rev.\ D {\bf 91}, 055026 (2015)
  [arXiv:1502.00919 [hep-ph]].  \\
  G.~M.~Pelaggi, A.~Strumia and S.~Vignali,
  ``Totally asymptotically free trinification,''
  JHEP {\bf 1508}, 130 (2015)
  [arXiv:1507.06848 [hep-ph]]. \\
  J.~E.~Camargo-Molina, A.~P.~Morais, A.~Ordell, R.~Pasechnik, M.~O.~P.~Sampaio and J.~Wessén,
  ``Reviving trinification models through an E6 -extended supersymmetric GUT,''
  Phys.\ Rev.\ D {\bf 95}, no. 7, 075031 (2017)
  [arXiv:1610.03642 [hep-ph]].
  
\bibitem{Li:1973mq} 
  L.~F.~Li,
  ``Group Theory of the Spontaneously Broken Gauge Symmetries,''
  Phys.\ Rev.\ D {\bf 9}, 1723 (1974).
  
\bibitem{Camargo-Molina:2013qva} 
  J.~E.~Camargo-Molina, B.~O'Leary, W.~Porod and F.~Staub,
  ``$\mathbf{Vevacious}$: A tool for finding the global minima of one-loop effective potentials with many scalars,''
  Eur.\ Phys.\ J.\ C {\bf 73}, no. 10, 2588 (2013)
  [arXiv:1307.1477 [hep-ph]].
  
\bibitem{Martin:2017lqn} 
  S.~P.~Martin,
  ``Effective potential at three loops,''
  arXiv:1709.02397 [hep-ph].
  
\bibitem{Vafa:1983tf} 
  C.~Vafa and E.~Witten,
  Nucl.\ Phys.\ B {\bf 234}, 173 (1984).
  
\bibitem{Ramsey-Musolf:2017tgh} 
  M.~J.~Ramsey-Musolf, P.~Winslow and G.~White,
  ``Color Breaking Baryogenesis,''
  arXiv:1708.07511 [hep-ph].
  
\end{thebibliography}
\end{document}